\documentclass[reprint,aps,prl,10pt]{revtex4-1}
\usepackage{graphicx,amsmath,amssymb,verbatim}
\pdfoutput=1

\begin{document}
\title{Design and Fabrication of a Chip-based Continuous-wave Atom Laser}
\author{E. P. Power}
\affiliation{Department of Physics, University of Michigan, Ann Arbor, MI 48109, USA}
\author{L. George}
\author{B. Vanderelzen}
\author{P. Herrera-Fierro}
\affiliation{Lurie Nanofabrication Facility, University of Michigan, Ann Arbor, MI 48109, USA}
\author{R. Murphy}
\affiliation{Applied Physics Program, University of Michigan, Ann Arbor, MI 48109, USA}
\author{S. M. Yalisove}
\affiliation{Department of Materials Science and Engineering, University of Michigan, Ann Arbor, MI 48109, USA}
\author{G. Raithel}
\affiliation{Department of Physics, University of Michigan, Ann Arbor, MI 48109, USA}

\begin{abstract}
We present a design for a continuous-wave (CW) atom laser on a chip and describe the process used to fabricate the device. Our design aims to integrate quadrupole magnetic guiding of ground state $^{87}\textrm{Rb}$ atoms with continuous surface adsorption evaporative cooling to create a continuous Bose-Einstein condensate; out-coupled atoms from the condensate should realize a CW atom laser. We choose a geometry with three wires embedded in a spiral pattern in a silicon subtrate. The guide features an integrated solenoid to mitigate spin-flip losses and provide a tailored longitudinal magnetic field. Our design also includes multiple options for atom interferometry: accomodations are in place for laser-generated atom Fabry-Perot and Mach-Zehnder interferometers, and a pair of atomic beam X-splitters is incorporated for an all-magnetic atom Mach-Zehnder setup. We demonstrate the techniques necessary to fabricate our device using existing micro- and nano-scale fabrication equipment, and discuss future options for modified designs and fabrication processes.
\end{abstract}
\maketitle

\section{Introduction}
\label{sec:intro}
The achievement of Bose-Einstein condensation (BEC) in cold atomic gases \cite{Anderson95,Davis95,Mewes96,Bradley97} has lead to a renewal in atomic physics and has opened an interface between atomic and condensed-matter physics. Investigations subsequent to the achievement of BEC have dealt with fundamental properties of BECs such as vibrational excitations of BECs \cite{Jin96,Mewes96b} and the role of the scattering length and its modifications by Feshbach resonances \cite{Inouye98,Courteille98}. BECs are often described as macroscopically occupied wave-functions. As such, they exhibit first-order coherence that leads to interference effects \cite{Andrews97} similar to those observed in optical interferometers. Unlike interfering laser beams, BECs are susceptible to phase shifts that are caused by the mean-field interaction contained in the Gross-Pitaevski equation \cite{Wallis98}. Thus, fringe shifts in matter-wave interference can arise from mean-field interactions with other atoms or BECs \cite{Myatt97,Hagley99}. Mean-field interactions also enable, for instance, non-linear four-wave-mixing of matter waves \cite{Deng99}. Further, unlike photons in laser beams, atoms in BECs are sensitive to DC and AC electromagnetic fields, gravitational fields, and rotations. These interactions also cause fringe shifts in matter-wave interference.

The high sensitivity of coherent matter waves to a wide class of physical phenomena~\cite{Andrews97,Wallis98,Myatt97,Hagley99} opens up new possibilities in high-precision sensing applications. In order to maximize the resolving power of such atom-interferometric sensors it will be necessary to operate them with continuous-wave (CW), phase- and amplitude-stable, coherent matter-wave fields (atom lasers). Atom lasers can be derived from BECs using various output coupling mechanisms. Pulsed atom lasers were realized soon after the first BEC demonstration by extracting BEC-atoms with radio-frequency (RF) \cite{Mewes97,Steck98,Bloch99} and Raman laser pulses \cite{Hagley99b,Debs:ramancoupler}; the Raman output coupling scheme was demonstrated to provide a higher continuous output flux than an RF-based coupler~\cite{debs:outcoupler}. The first-order \cite{Kohl01} and second-order \cite{Ottl05} temporal coherence properties of such sources have also been measured. In order to prepare a phase- and amplitude-stable CW atom laser, a phase- and amplitude-stable CW BEC will be required, first. A BEC can, in principle, be sustained forever by pumping it with a permanent, dense current of cold but still non-condensed atoms \cite{Holland96,Williams98}, which one may introduce by optical pumping \cite{Bhongale00,Santos01,Floegel03}. A step in this direction was made by Chikkatur {\sl et al.} \cite{Chikkatur:continuousBEC}, who achieved a CW BEC by pulsed re-loading with cold atoms transported by optical tweezers, and by Robins {\sl et al.} \cite{robins:recentatomlaser}, who used a combination of RF fields and optical pulses to fill the lasing BEC with atoms from another BEC prepared in a different hyperfine state. In related work, the merging of multiple condensates was studied experimentally \cite{Shin04} and in theory \cite{Yi05}. While these methods have successfully yielded a CW BEC, it seems difficult to develop them in a way that would also deliver phase- and amplitude-stability from one reloading cycle to the next. Phase- and amplitude-stability of a BEC will require the absence of any time-dependent trapping fields in the vicinity of the BEC. This critical condition precludes the application of time-dependent laser, magnetic, and RF fields to the atoms that are to be condensed.

A promising avenue for preparing phase- and amplitude-stable CW BECs and atom lasers is to continuously laser-cool, evaporatively cool and condense an atomic flow in a guided atomic beam architecture. In this arrangement laser-beam guides \cite{Schiffer98,Kuppens98,Torii98,Couvert:opticalguiding} and hollow-core fiber-optic guides \cite{Renn95,Muller00} will be of limited usefulness, since atoms guided in these guides are heated by photon scattering. Also, the range of geometries feasible with fibers is restricted; in the case of hollow-core fiber guides it can be difficult to achieve a satisfactory vacuum pressure. Magnetic atom guides are better suited because the guide losses and heating due to photon scattering can be eliminated and because a variety of geometries (variable field gradients, variable guide depth, curves in guides, unlimited length, etc.) can be realized. Magnetic atom guides have been developed by multiple groups using macroscopic wire structures \cite{Denschlag99,Teo01,Sauer01}. The wires were, in some cases, embedded in glass fibers \cite{Key00} and their magnetic fields amplified by ferromagnetic insets \cite{Vengalattore02,Rooijakkers03,Teo02a}. Multiple groups have built micro-fabricated magnetic atom guides \cite{Fortagh98,Folman00,Fortagh00,Hansel01,Drndic98,Dekker00,Muller99} and have implemented atomic-flow control elements such as switches \cite{Muller01} and splitters \cite{Cassettari00}, while others have used microscopic magnetized surfaces \cite{Hinds98}. Microfabricated magnetic guides have been combined with quasi-magnetostatic \cite{Hansel01} and electrostatic conveyor-belt potentials \cite{Kruger03} that can be used to promote a controlled atom flow. Clouds of cold atoms have also been transported using large-scale moving magnetic potentials \cite{Greiner01}.

A continuous, phase- and amplitude-stable, magnetically guided atom laser is ideal for the realization of atom-interferometric devices and sensors. Free-space atom interferometers have previously been implemented based on coherent beam-splitting and recombination by Raman transitions \cite{Gustavson97,Snadden98,McGuirk00}, matter-wave gratings \cite{Chapman95,Chapman95b,Ekstrom95,Schmiedmayer95} and standing-wave light fields \cite{Rasel95}. These devices were used to measure rotation in a Sagnac geometry \cite{Gustavson97}, gravity \cite{Snadden98,McGuirk00,Dimopoulos:gravimeter}, atom-light interactions \cite{Chapman95,Rasel95}, the index of refraction of low-density gases \cite{Chapman95b,Schmiedmayer95}, and electric polarizability \cite{Ekstrom95}. More recently, atom interferometers fabricated on chips have been studied \cite{Cassettari00b,Andersson00,Brugger05,Wang05}. It has been found, however, that matter-waves close to surfaces experience considerable degradation effects in homogeneity and phase \cite{Hinds01,Leanhardt02,Dettmer02}. Therefore, guide-based interferometers will likely require guide diameters that are not too small (of order tens to hundreds of $\mu$m).

\section{Design}
\label{sec:design}
Our scheme relies on the use of surface adsorption by a dielectric surface to evaporatively cool the atoms to degeneracy. BEC formation was demonstrated using this technique~\cite{harber:surfaceadsorption}, however the apparatus used did not lend itself to continuous loading or out-coupling. Using a magnetic atom guide we can overcome this limitation by loading packets or a continuous stream of atoms at one end of the guide and implementing distributed evaporative cooling along the guide length by gradually moving the guide towards the adsorption surface. This scheme, however, is challenging in macroscopic guide designs. Most guides considered for this purpose are several meters in length, which would necessitate installing a dielectric cooling surface with a flatness on the order of $10~\mu\textrm{m}$ over the length of the guide. Furthermore, potential applications using these large guides are limited due to their unwieldy size and requirement for power supplies producing hundreds of amperes.

\begin{figure}[t]
    \begin{center}
    \includegraphics[bb=184 323 424 467,width=3.375in]{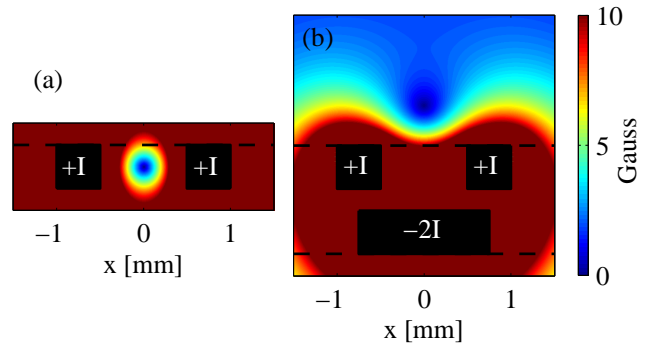}
        \end{center}
    \caption{(a) Magnetic field calculation for a two-wire design. (b) Magnetic field for our three-wire design. The location of the chip surfaces are indicated by dashed lines, and the wire locations are marked as black patches. Ground state $^{87}\textrm{Rb}$ atoms in a low magnetic field seeking state are trapped in the field minimum.}
    \label{fig:2vs3wires}
\end{figure}

Instead, we choose to adopt a chip-based guide, where the wires producing the guiding potential are embedded in a 100~mm diameter silicon wafer and spiral towards the center of the chip. A previous demonstration of such a device~\cite{Luo:spiralguide} lacked the necessary length to cool the atoms to degeneracy. Calculations indicate that a guide length of $\sim 50~\textrm{cm}$ is sufficient~\cite{dalibard:condensation} for continuous surface adsorption to cool atoms to a phase space density $n_0\lambda^3 \geq 2.612$, the critical density for condensation, given reasonable input parameters: $100~\mu\textrm{K}$ initial temperature and a loading rate of $3\times 10^9\textrm{s}^{-1}$. Our particle-in-cell simulations show similar results~\cite{Hempel:cwatomlaser}. Our design implements a 3-wire geometry to produce the guiding potential, as shown in Fig.~\ref{fig:2vs3wires}. Located on the underside of the chip, the third wire carries the sum current of the two guide wires in the anti-parallel direction, producing an additional horizontal field component and pushing the center of the guiding channel above the chip surface.

\begin{figure}[t]
    \begin{center}
    \includegraphics[bb=0 0 244 218,width=3.375in]{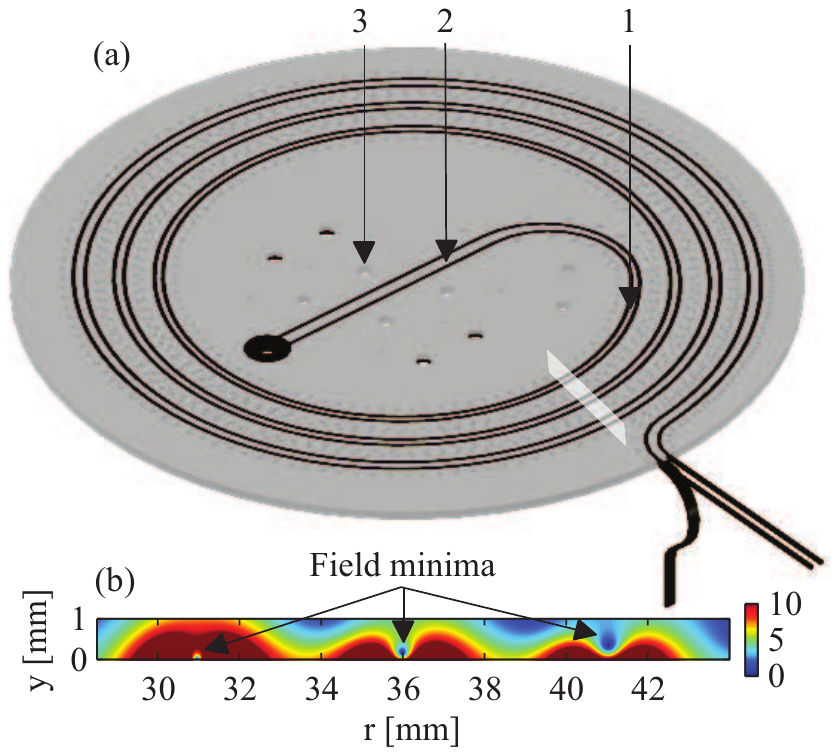}
        \end{center}
    \caption{(a) Top-view rendering of the atom laser chip. Arrow 1 points to the hole for the far off-resonant laser used as a light-shift barrier, arrow 2 points to the slot for the two Fabry-Perot beams, and arrow 3 points to one of the mirror mount holes for the Mach-Zehnder setup. (b) 3D magnetic field calculation in the plane indicated by the semi-trasparent rectangle in (a). The guide wire separation decreases from 1.6~mm to 1~mm along the first $\sim70~\textrm{cm}$ of the guide, lowering the guide center closer to the chip to implement continuous distributed surface adsorption evaporative cooling. As the channel height decreases, the transverse gradient increases, providing the magnetic compression necessary to increase the collision rate (required for efficient evaporative cooling).}
    \label{fig:chiptopandfield}
\end{figure}

A modest current $I=5~\textrm{A}$ in each guide wire and $I=-10~\textrm{A}$ in the bias wire is sufficient to produce a guiding channel with transverse gradient $|\textrm{dB/dr}|=102~\textrm{G/cm}$, located $465~\mu\textrm{m}$ above the chip. The total power dissipation is $<5~\textrm{W}$. The guide wires have a square cross-section with 0.5~mm width and initial center to center separation of 1.6~mm. The bias wire on the bottom of the chip has dimensions $1.5\times0.5~\textrm{mm}$ and has a center located 0.75~mm below the guide wire center. The guide follows three full revolutions around an Archimedean spiral with a change in radius of $-5~\textrm{mm}/2\pi$, yielding a guide length of $\sim70~\textrm{cm}$, as shown in Fig.~\ref{fig:chiptopandfield}(a).

Our evaporation ramp is implemented by gradually reducing the separation between the guide wires from 1.6~mm to 1~mm over the 70~cm guide length. As shown in Fig.~\ref{fig:chiptopandfield}(b), this gradually reduces the height of the guiding channel while simultaneously providing magnetic compression (the final height of the guiding channel is $25~\mu\textrm{m}$). After 70~cm the transverse field gradient is $|\textrm{dB/dr}|=929~\textrm{G/cm}$. Global height variations in the guiding channel due to manufacturing imperfections are compensated through the use of a center tap at the location where the guide wires and bias wires meet; control of the channel height is achieved by adjusting the current ratio between the guide wires and bias wire. Following the 70~cm cooling zone the guide wires remain at a fixed separation for an additional $\sim30^{\circ}$ turn around the spiral, at which point a $300~\mu\textrm{m}$ diameter hole is placed through the chip and bias wire. This hole is to accommodate a far off-resonant laser to close the trap and form a three-dimensional potential minimum for condensation. Tuning the intensity of the barrier to create a non-zero tunneling rate extracts atoms from the BEC and propagates this beam, a CW atom laser, to the interferometry area located in the straight guide segment at the center of the chip. As atoms move around the corner from the light-induced tunelling barrier to the interferometry area the guide wire separation increases to 2~mm, raising the guide height to $950~\mu\textrm{m}$ and decreasing the transverse gradient to $29~\textrm{G/cm}$.

\begin{figure}
    \begin{center}
    \includegraphics[bb=184 251 427 539,width=3.375in]{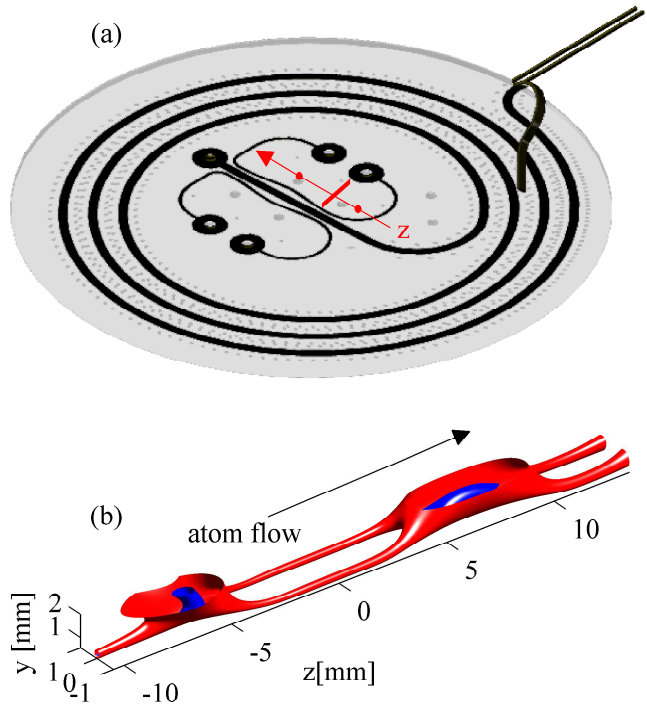}
        \end{center}
    \caption{(a) Bottom-view rendering of the atom laser chip. Two additional wires on the bottom layer are used to create a pair of magnetic hexapole beam splitters. (b) Isopotential surfaces calculated through the all-magnetic interferometer: $\textrm{red}\rightarrow|\textbf{B}|=8\textrm{G}$, and $\textrm{blue}\rightarrow|\textbf{B}|=2\textrm{G}$. The arrows in (a) and (b) indicate the direction of atomic flow. A cross mark through the arrow in (a) indicates the $z=0$ position in (b). At $z\approx-6~\textrm{mm}$ a second quadrupole channel merges with the guide channel to form a nearly pure hexapole potential, after which the two quadrupole channels split in a side-by-side configuration. At $z\approx7~\textrm{mm}$ the channels merge in a second hexapole beam splitter and then split again into parallel channels for readout. In (a), the locations of the two hexapole beam splitters are indicated by dots on the z-axis arrow.}
    \label{fig:chipbottomandint}
\end{figure}

Our design includes several options for atom interferometry experiments. A slot with $300~\mu\textrm{m}$ width and 2~mm length is cut through the chip and bias wire almost exactly at the center of the chip; two off-resonant beams through this slot will be used to set up an atom Fabry-Perot~\cite{wilkens:fabryperot}. Two 5~mm diameter holes on each side of the guide wires accommodate mirror mounts; laser beams parallel to the chip will use these mirrors to form standing-wave gratings for a Mach-Zehnder interferometer. The third option is shown in Fig.~\ref{fig:chipbottomandint}; two butterfly wing shaped wires on the bottom of the chip each carry $\sim1~\textrm{A}$ in the same direction as the bias wire and modify the trapping potential to create two regions where the quadrupole term vanishes, leaving only a hexapole magnetic topology. As seen in Fig.~\ref{fig:chipbottomandint}(b), at the first hexapole beam splitter a second quadrupole channel comes down from above the chip to merge with the guide channel. This occurs a few millimeters prior to the hexapole wires' closest approach to the bias wire. As the hexapole wires move closer to the bias wire, the two quadrupole channels split into a side-by-side configuration, merging again $\sim13~\textrm{mm}$ later in a second hexapole beam splitter which forms the analyzer for this interferometer. The two channels are again split side-by-side for easy probe access to both output channels.

\begin{figure}[t]
    \begin{center}
    \includegraphics[bb=184 310 427 481,width=3.375in]{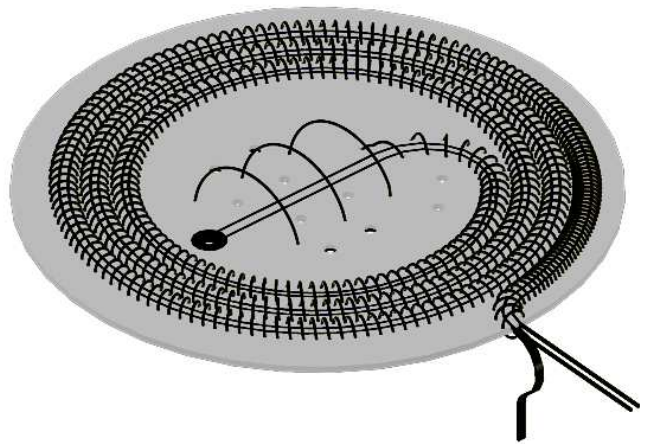}
        \end{center}
    \caption{Top-view rendering of the atom laser chip. The solenoid sewn through the chip prevents Majorana spin-flip losses in the guide, provides a shallow longitudinal trap preventing some backwards-moving atoms from leaving, and directly counters the uphill gravitational potential present as atoms move from the BEC into the interferometer zone.}
    \label{fig:chiptopandsolenoid}
\end{figure}

As the atoms' temperature drops towards that required for condensation, mitigating Majorana spin-flip losses~\cite{sukumar:spinflip,brink:spinflip} due to the magnetic field zero at the center of the guiding channel becomes paramount. In a linear guide geometry a solenoid co-axial with the guide axis produces a constant longitudinal magnetic field, eliminating the field zero at the trap center; this longitudinal offset is critical to obtaining a condensate. The previously demonstrated spiral guide~\cite{Luo:spiralguide} did not include a mechanism for producing an offset to eliminate the field zero, and thus is incapable of reaching BEC conditions. Producing a magnetic field offset could be accomplished by running a high current through a large wire straight through the chip center. However, as atoms spiral inward the strength of the offset field would increase, producing an uphill slope and potentially defeating our guide. Instead, we choose to use a co-axial solenoid as shown in Fig.~\ref{fig:chiptopandsolenoid}.

The solenoid has an initial pitch of 10 turns/cm and a diameter of 5~mm; for a wire carrying $620~\textrm{mA}$ this produces a longitudinal field of $7.8~\textrm{G}$. As one would expect, the gradual spiral bend in the solenoid does not significantly impact its performance. 3D calculations demonstrate that the magnetic field inside the solenoid differs from the ideal $B=\mu_0NI/l$ by less than 0.1\%, even in the presense of variations in loop diameter of up to 10\%. At the center of the guide, these small variations correspond to fluctuations in the trapping potential of only $\sim2~\textrm{kHz}$. After traveling $\sim90^{\circ}$ around the spiral the pitch reduces to 5.1 turns/cm, reducing the longitudinal offset field to $4~\textrm{G}$. The shallow drop in the longitudinal component of the guiding potential promotes the flow of atoms through the cooling region and into the BEC zone of the chip. In the BEC zone, the solenoid maintains its 5.1 turns/cm pitch until just beyond the hole for the far off-resonant barrier laser, at which point the pitch decreases again to 0.96 turns/cm and the diameter increases to 2~cm. This last reduction in pitch is designed to directly counter the uphill gravitational potential created when the guide wire separation increases from 1~mm to 2~mm in the interferometer area. The increase in channel height of nearly 1~mm is directly countered by a downhill magnetic potential to prevent atoms from getting stuck immediately after tunneling through the barrier laser. After this final pitch decrease the solenoid produces an offset of $0.75~\textrm{G}$, which will still be sufficient to prevent spin-flip losses.

\section{Loading the Guide}
\label{sec:loading}
The entrance turn into the spiral has a radius of 6~mm and can guide atoms moving at speeds up to $62~\textrm{cm/s}$. For our initial experiments we will capture atoms from the output of a Zeeman slower~\cite{Phillips:Zeeman}, which will inject a flux of $\sim3\times10^{11}\textrm{s}^{-1}$ of slow atoms into a magneto-optical trap (MOT). Atoms trapped in this MOT are then transferred via a pusher beam to a second MOT located between two macroscopic guide wires, each carrying $I=65~\textrm{A}$. Separated by 17~mm, the 65~A wires are angled downward relative to the chip plane at an angle of $10^{\circ}$. Once the second MOT is full the secondary MOT coils are turned off, allowing the magnetically trapped atoms to slide down towards the chip between the 65~A wires. The atoms enter the chip with a forward velocity of $44~\textrm{cm/s}$. The 65 A wires are coupled to off-chip stubs from the chip guide wires $\sim14~\textrm{cm}$ from the chip; 60 A of the 65 A in each guide wire is diverted, leaving only 5 A on each guide wire as the atoms move toward the chip. In this region the guide wire separation also tapers from 17 mm to 1.6 mm over a 4 cm distance, providing adiabatic magnetic compression between the MOT field gradient and the guide's initial field gradient.

For initial demonstrations of magnetic guiding capability a pulsed scheme will be used: atoms are prepared in the secondary MOT, released onto the chip guide, and probed at various locations along the guide spiral. The pulsed scheme will also allow us to characterize evaporative cooling via surface adsorption on the chip's bare silicon surface. In this mode, the atoms are guided in the low-field-seeking state $|F=2,m_F=2>$, which has a magnetic moment of one Bohr magneton. Since magnetically guided atoms are susceptible to photon scattering, all laser cooling light must be turned off while the guiding is in progress. For subsequent studies towards CW operation, a scheme will be needed that is compatible with simultaneous operation of the magnetic guide with the secondary MOT.
Stray light from the lasers employed in the MOT has the potential to reach magnetically guided atoms on the chip. A total loss of guided atoms in the state $|F=2,m_F=2>$ would be virtually guaranteed. A well-proven method to prevent this is to optically pump the atoms into the $|F=1,m_F=-1>$ state as they enter the chip~\cite{teo:depumper}. This state has a magnetic moment of 0.5 Bohr magnetons and is also magnetically guided, but is not resonant with MOT laser cooling light.

\section{Fabrication}
\label{sec:fab}
Now that we have a reasonable design, our objective is to develop a fabrication process that is robust and reproducible. Integrating a magnetic atom guide onto a silicon wafer allows us to leverage existing micro- and nano-fabrication technologies designed for micro-electro-mechanical systems (MEMS) and semiconductor electronics. The photolighographic, thin film growth and deposition, and etching tools available in our clean room are well characterized and backed by a wealth of experienced use. The features on our atom chip (wire channels, through holes, etc.) are far larger than the mircometer-scale features usually created with these tools, however with some experimentation we show that it is possible to use exisiting clean room techniques to fabricate our device.

\begin{figure}[t]
    \begin{center}
    \includegraphics[bb=184 273 427 517,width=3.375in]{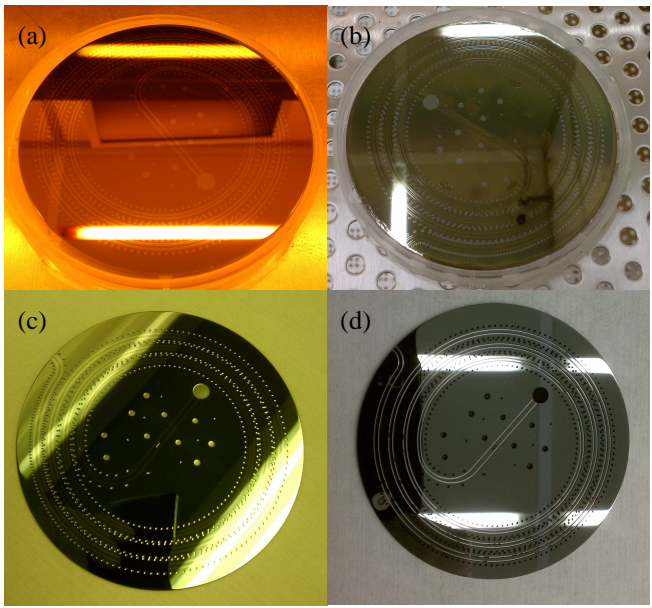}
        \end{center}
    \caption{(a) Wafer with top-side mask with all device features exposed. (b) Wafer with a second top-side mask containing only holes, protecting the guide wire channels during the first etch. (c) Wafer following the initial etch of holes only to a depth of $300~\mu\textrm{m}$. (d) Wafer after the $2^{\textrm{nd}}$ etch: holes are $850~\mu\textrm{m}$ deep and the guide wire channels are $550~\mu\textrm{m}$ deep.}
    \label{fig:wafermasking}
\end{figure}

Fabrication of the atom laser chip begins by placing a 1.25~mm-thick, double-side polished Si wafer with no flats in a furnace: a wet oxidation process $(\textrm{Si}+2\textrm{H}_2\textrm{O}\rightarrow \textrm{SiO}_2 + 2\textrm{H}_2)$ at $1100^{\circ}$~C is used to grow a $1.5~\mu\textrm{m}$-thick thermal oxide layer on the wafer. A $3~\mu\textrm{m}$-thick layer of $\textrm{SPR}^{\textrm{TM}}220$ photoresist is applied using a spin-coater and soft baked, after which our first lithography mask is exposed and developed, as shown in Fig.~\ref{fig:wafermasking}(a). Our first mask contains all device features on the top side of the wafer: guide wire channels, holes for the solenoid, holes for far off-resonant lasers, and mounting and screw holes. A wet etch is performed in a bath of buffered hydrofluoric acid (BHF) to remove the oxide exposed by the mask, creating an oxide hard mask copy of the original. The photoresist is then removed and a $4~\mu\textrm{m}$-thick layer of $\textrm{SPR}^{\textrm{TM}}220$ is spun on. A second mask is exposed; the second mask contains all the features of the first mask except for the guide wire channels. The chip now has two masks: the oxide mask containing all top layer features which is covered by a photoresist mask containing just the features that must be etched through the chip, as shown in Fig.~\ref{fig:wafermasking}(b).

A deep reactive ion etch (DRIE) is performed using the Bosch process~\cite{Laermer:Bosch} to etch the exposed Si to a depth of $300~\mu\textrm{m}$, as shown in Fig.~\ref{fig:wafermasking}(c). The Bosch process alternates between depositing $\textrm{C}_4\textrm{F}_8$ and bombarding the wafer with a mixed-species plasma composed of oxygen and fluorine dissocated from $\textrm{SF}_6$ in a high power RF field. The $\textrm{C}_4\textrm{F}_8$ acts as a passivation layer, preventing the F from reacting with the Si. The plasma is highly directional; ballistic ions, primarily $\textrm{SF}_5^+$, break through the $\textrm{C}_4\textrm{F}_8$ layer at the bottom of features to be etched, leaving the passivation in place on the side walls. Once the passivation is removed the exposed Si reacts with neutral F to produce $\textrm{SiF}_4$~\cite{garrison:DRIE}, which desorbs from the surface and flows out through the turbo pump attached to the etching chamber. The $\textrm{C}_4\textrm{F}_8$ passivation on the side walls results in a highly anisotropic etch process and features that are etched almost straight down.

After etching the through-holes on the chip, the photoresist mask is stripped, and the holes and guide wire channels are etched together until the guide wire channels are $550~\mu\textrm{m}$ deep and the holes are $850~\mu\textrm{m}$ deep, as shown in Fig.~\ref{fig:wafermasking}(d). Our etch recipe had to be modified somewhat from the standard all-purpose recipe, since the percentage of exposed Si on the chip $(\sim20\%)$ is much higher than is usual for this process. The high percentage of exposed Si area leads to fluorine starvation as all available F atoms are consumed; this depletion of the etchant causes etch rate variations in excess of 10\% when measuring the etch depth in areas with large amounts of exposed Si vs. areas with little exposed Si. Such a variation is problematic, since our design cannot accommodate over-etching by this amount; when too much material is removed the top-side and bottom-side wire channels do not have enough Si between them for the chip to maintain mechanical integrity, and it falls apart under its own weight. To combat this issue we doubled the $\textrm{SF}_6$ flow rate from the standard recipe, decreased the platen temperature to mitigate the extra heating encountered when reacting large quantities of Si during each etch cycle, and adopted a two-phase etch process. After completing half of the etch, the wafer is removed from the chamber, rotated by $180^{\circ}$ on its carrier wafer, and etched for the remaining $\sim275~\mu\textrm{m}$. We found that even with the larger $\textrm{SF}_6$ flow rate a linear variation in etch rate was present between the side of the wafer closest to the turbo pump and the opposite edge. Rotating halfway through the etch was the only solution we tried that evened out the etch rate and produced acceptable results. With our modified recipe we etched to an average depth of $552~\mu\textrm{m}\pm2\%$.

\begin{figure}
    \begin{center}
    \includegraphics[bb=184 334 427 456,width=3.375in]{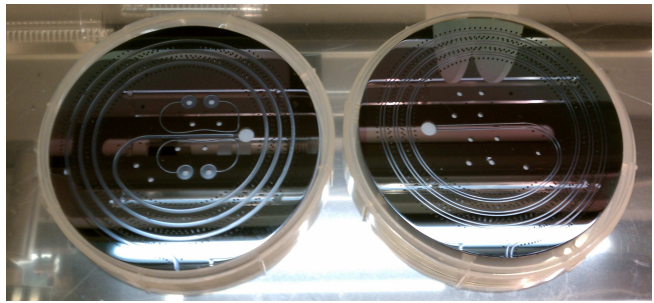}
        \end{center}
    \caption{Atom laser chip after the final DRIE process.}
    \label{fig:finished_substrate}
\end{figure}

Once the guide wire channels are complete the remaining oxide mask is stripped in another BHF bath; the wafer is then returned to the furnace to grow another $1.5~\mu\textrm{m}$-thick layer of thermal oxide. The large number and extreme depth of the features on the top side of the wafer make it impossible for a vacuum chuck to hold this surface. In order to process the bottom layer, we mount the top surface of the chip to a glass carrier wafer using a spun-on mixture of $\textrm{Crystabond}^{\textrm{TM}}$ 509 with acetone, 20:80 by weight. The spun-on $\textrm{Crystabond}^{\textrm{TM}}$ is transparent and allows us to see the alignment marks on the top surface quite easily. A $3~\mu\textrm{m}$-thick layer of $\textrm{SPR}^{\textrm{TM}}220$ is spun onto the bottom surface of the chip, and a single mask containing all features on the underside of the chip is aligned, exposed, and developed. The exposed oxide is removed in the BHF bath, after which the chip is removed from the glass carrier by soaking it in acetone, which also removes the photoresist. After removing any residual photoresist we etch the bottom layer using the DRIE tool to a depth of $550~\mu\textrm{m}$. The holes through the chip connect from the back side during this etch. The finished chip, shown in Fig.~\ref{fig:finished_substrate}, is returned to the furnace for a final $1.5~\mu\textrm{m}$-thick layer of thermal oxide, which is needed to prevent the copper wires from shorting to the chip.

\begin{figure}
    \begin{center}
    \includegraphics[bb=184 273 427 517,width=3.375in]{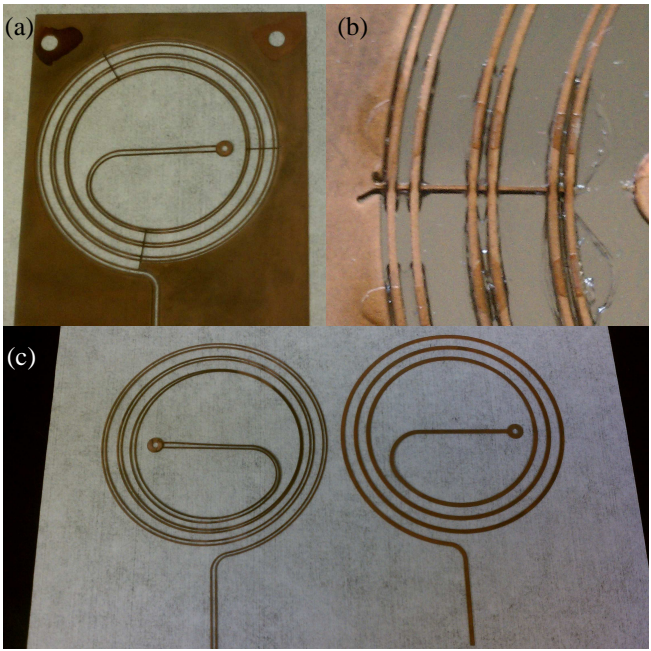}
        \end{center}
    \caption{(a) Guide wires were fabricated with three support ribs to ensure machining accuracy. (b) Femtosecond laser ablation was used to cut the ribs away from the wires. (c) Finished guide (left) and bias (right) wires.}
    \label{fig:wires}
\end{figure}

The copper wires for the guide wire, bias wire, and hexapole beam splitter wires are fabricated from oxygen-free electronic (OFE) grade sheet copper using a wire electrical discharge machining (EDM)~\cite{Todd:wireedm}. The sheet copper is sandwiched between two aluminum plates during the EDM to prevent it from springing out of position or curling away from the chuck. Even so, our machinist found it impossible to cut the complete spiral shape of either the guide wires or the bias wire without leaving some support ribs in place. Three $\sim100~\mu\textrm{m}$ wide ribs spaced $120^{\circ}$ apart were left in place to provide the stability required to complete the machining, as shown in Fig.~\ref{fig:wires}(a). Mechanical removal of the ribs would be extremely difficult to accomplish without causing damage to the wires; in some locations the rib joins two wire segments that are spaced only 0.5~mm apart. Instead, we opted to use femtosecond laser ablation to remove the supports.

Machining was performed using a Clark-MXR CPA-2001 Ti:sapphire ultrafast femtosecond ($1~\textrm{fs}=10^{-15}~\textrm{s}$) laser with a 150 fs pulse duration, central wavelength of 780 nm, repetition rate of 1 kHz, maximum pulse energy of 1 mJ, and a Gaussian intensity profile. Femtosecond laser damage of metals and semiconductors is a uniquely deterministic process vs. laser machining using longer pulse lengths and has been extensively studied~\cite{ablation:1,ablation:2,ablation:3,ablation:4,ablation:5,ablation:6}. The large electric field of each laser pulse excites electrons to high electronic states; thermal re-equilibration with the cold lattice leads to complete melting of the first few nanometers of material. Subsequent rapid thermal expansion of this melt leads to mechanical failure and ejection of a molten slab of material from the bulk. In this fashion, machining of a metal can be accomplished with minimal collateral damage to the surrounding area. To machine the $500~\mu\textrm{m}$ thick support ribs for the copper guide and bias wires we mount the wires to a carrier wafer, as shown in Fig.~\ref{fig:wires}(b), and mount this assembly to a 4-axis translation stage which translates the wires in front of the 1 kHz laser pulse train. The beam is focused with a $f=20~\textrm{cm}$ lens to a diameter of $\sim40~\mu\textrm{m}$ and a peak fluence of $8~\textrm{J/cm}^2\textrm{-pulse}$. The stage is translated at a velocity of 1 mm/s in front of the focused beam, yielding a pulse-to-pulse overlap of 98\%.  After 10 passes over the same area the sample is moved $250~\mu\textrm{m}$ towards the lens to advance the focal spot deeper into the copper support rib. Advancement of the sample towards the lens is performed twice; a total of 30 passes through the laser is required to cut completely through the ribs. With this technique we were able to remove the supports to within $10~\mu\textrm{m}$ of the idealized wire shape. After dismounting from the carrier wafer a quick touch-up with a diamond file removes any excess Cu left over from the laser machining. The finished wires, shown in Fig.~\ref{fig:wires}(c), are now within design specifications and ready for oxide removal and mounting.

A critical aspect of our design is that the center of mass of each wire remain a constant distance from the surface of the chip to prevent the guide channel from undulating up and down. The sheet copper used to fabricate the wires was purchased overly-thick and then lapped to the target thickness, guaranteeing us a set of wires with minimal thickness variations. With wires such as these we can safely register the top surface of the wires to the chip surface and be confident that the wire center of mass is always at a constant distance. We fabricated several additional wafers to assist us with the wire mounting. First, we anodically bond a $500~\mu\textrm{m}$-thick Si wafer to a glass wafer and then etch the guide wire channels in the Si wafer using the DRIE tool, etching clear through to the glass. We make a similar piece for the wires on the bottom of the chip. Next, we lap away $\sim200~\mu\textrm{m}$ of Si from both of these wire formers, leaving wire channels shallower than the thickness of the wires. Separately, we fabricate two handle wafers, one for each side of the chip: the handles contain the reverse image of the wire channels, but narrower ($100~\mu\textrm{m}$ width instead of $550~\mu\textrm{m}$). DRIE is used to remove $\sim250~\mu\textrm{m}$ of Si from the handle wafers, leaving the wire patterns in bas relief. Eight symmetrically placed 1~mm-diameter pedestals are also left on the handle wafers to allow the atom laser chip surface to register with the wire surface.

\begin{figure}[t]
    \begin{center}
    \includegraphics[bb=184 334 427 456,width=3.375in]{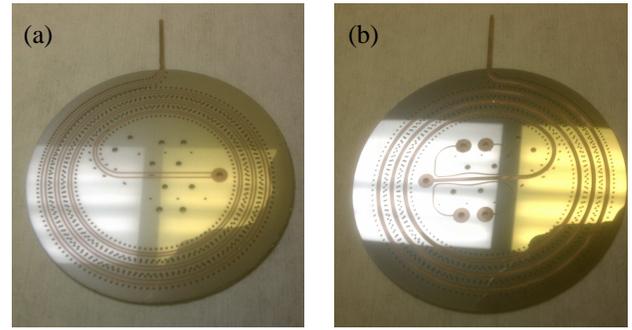}
        \end{center}
    \caption{(a) Top side of the atom laser chip after polishing. (b) Bottom side after polishing.}
    \label{fig:polishedchip}
\end{figure}

A heat-curable acetone-soluble temporary mounting adhesive is spun on to the handle wafers at 5~krpm to produce a $\textrm{sub-}\mu\textrm{m}$-thick film of adhesive. The wires are acid etched to remove any oxide and then placed in their respective wire formers, and the handle wafers are inverted and placed on top, with the thin wire-shaped pedestals exactly centered on each wire using alignment pins built into the former assemblies and handle wafers. The adhesive in the wafer-wire-wafer sandwich is heat cured on a hot plate, using a heavy weight to ensure the handles remain firmly in contact with the wires. Once the adhesive is cured the handles are removed, taking the copper wires with them. A UHV-compatible, thermally conductive, electrically insulating epoxy is injected into the atom chip wire channels using a syringe, and the handle-mounted wires are then inserted into the channels. When the wires are fully inserted into the atom chip the eight 1~mm-diameter pedestals located on the handle wafers contact the surface of the atom chip and prevent any further insertion. The UHV epoxy is cured, and the handle is removed by soaking in acetone.

\begin{figure}[t]
    \begin{center}
    \includegraphics[bb=184 251 427 539,width=3.375in]{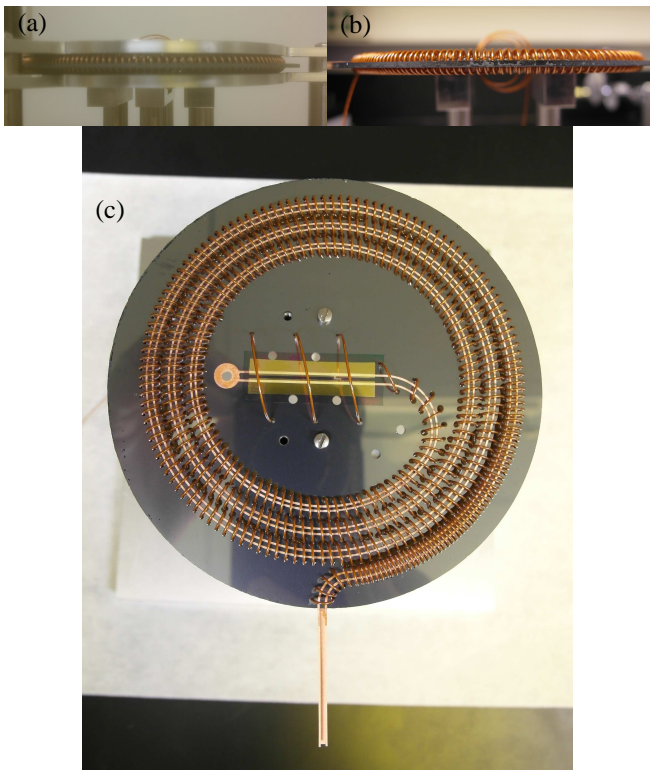}
        \end{center}
    \caption{(a) A special jig is used to set the diameter of each turn of the solenoid to 5~mm. (b) Chip and solenoid after sewing. (c) Top view, ready for mounting. }
    \label{fig:chipsolenoid}
\end{figure}

Excess epoxy is removed by polishing using a slurry of $1~\mu\textrm{m}$ colloidal alumina in deionized water; this process also removes the surface layer of thermal oxide, as shown in Fig.~\ref{fig:polishedchip}, leaving a bare Si surface for evaporative cooling of the atoms. Two gold mirrors were added in the interferometer zone of the chip to be used for shadowgraphy, should it be experimentally necessary. The mirrors are electrically insulated from the chip and each other, allowing us to independently adjust the potential on each gold patch. A mask is made from Katpon\circledR~tape, and a $2000~{\AA}$-thick layer of $\textrm{SiO}_2$ is sputtered on, followed by a $200~{\AA}$-thick layer of Ti, a $\sim2~\mu\textrm{m}$-thick layer of Au, and a $2970~{\AA}$-thick layer of $\textrm{SiO}_2$ ($\lambda/4$ thickness at 780~nm).

The final step in assembling the atom laser chip is to sew the solenoid. We accomplish this using a sewing jig that suspends the wafer in mid-air between two precisely positioned aluminum plates, spaced 5~mm apart, as shown in Fig.~\ref{fig:chipsolenoid}(a). For each solenoid hole the alignment plates are removed, the Kapton\circledR-dipped 30 AWG wire is pulled through until the half-turn under consideration looks approximately correct, and the alignment plates are re-installed. The wire is adjusted using tweezers until it just touches the top and bottom alignment plates on each turn, with the result being a near-constant diameter, as seen in Fig.~\ref{fig:chipsolenoid}(b). Once the solenoid is complete it is epoxied in place using the same UHV-compatible epoxy used to hold the guide/bias/hexapole wires, after which the atom laser chip, seen in Fig.~\ref{fig:chipsolenoid}(c), is complete.

\begin{figure}[t]
    \begin{center}
    \includegraphics[bb=184 274 427 517,width=3.375in]{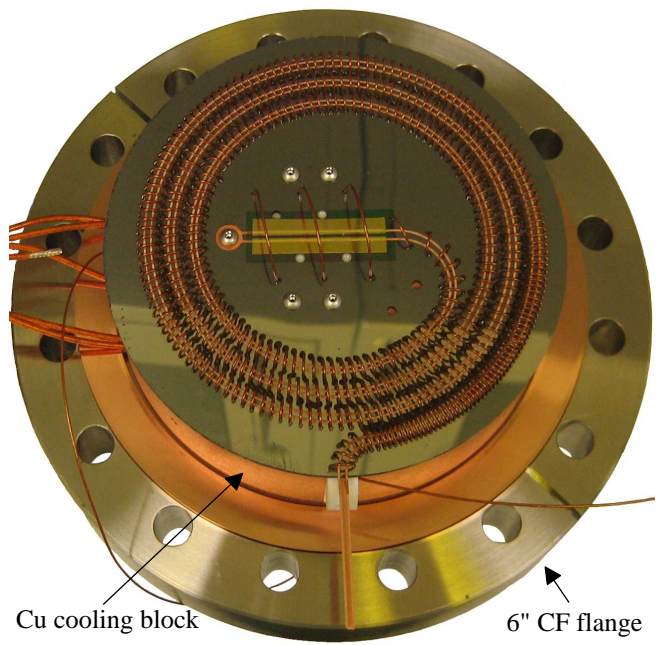}
        \end{center}
    \caption{Atom laser chip ready for installation. }
    \label{fig:chipmounted}
\end{figure}

The atom laser chip is mounted on an OFE copper heat sink as shown in Fig.~\ref{fig:chipmounted}, which in turn is mounted on a water-cooled 6" CF flange. The chip produces $<5~\textrm{W}$ of heat, which can be easily removed with our mounting setup. The Cu cooling block incorporates a spiral groove for the bottom-side portion of the solenoid, electrically insulated feedthrus for the center tap and hexapole wires, and optical access for the off-resonant laser beams. Prior to mounting we acid etch both the block and the chip one final time to remove any copper oxide. Once the copper is clean, we apply a thick bead of a more viscous UHV-compatible epoxy to the spiral groove of the cooling block; this epoxy effectively pots the underside of the chip to the block, improving thermal conduction away from the spiral wires. Small pieces of Kapton\circledR~sheet are rolled into cones and placed under both holes for the far off-resonant lasers to prevent epoxy from blocking our optical access. Fig.~\ref{fig:chipmounted} shows the final product, ready for installation and testing.

\section{Conclusion and Discussion}
Our design for a chip-based CW atom laser presents many fabrication hurdles, however we have demonstrated that it is possible to use micro- and nano-scale fabrication tools to realize the device. We incorporate only proven technologies in our device; although the integration of surface adsorption evaporation and magnetic atom guiding has never been demonstrated, we believe our chip stands a good chance of proving this blend of technologies. As of this writing, the chip is installed in its chamber and setup of the MOT and chip imaging optics is ongoing. Once completed, our first tasks will be to show guided atoms all the way to the center of the chip, followed by a demonstration of surface adsorption evaporative cooling. After reaching these milestones we can move forward with creating a BEC and setting up the atom Fabry-Perot to characterize the atom laser output.

Of the many steps involved in fabricating the atom laser chip, perhaps the most troublesome is the DRIE processing. Variation in the mounting of the chip to its 6"-diameter carrier wafer, contamination of the etch tool from other users' processes, and temperment of the DRIE tool all lead to a less than ideal etch success rate: only two of the ten wafers we started with survived through the final etch. A major improvement could be made by purchasing Si wafers cut with the [110] direction normal to the surface and replacing the DRIE process with a timed KOH wet etch. In such a scheme the etch rate will be perfectly uniform across the device, eliminating our reliance on an extremely complex and fickle DRIE tool to properly perform its function. We chose [100]-cut wafers for cost saving reasons, however hindsight shows us that the added expense of [110] wafers is more than offset by the elimination of the DRIE tool from our process.

Finally, the chip may be modified to accomodate MEMS-based light shutters. As mentioned in Sec.~\ref{sec:loading}, stray light penetrating from the MOT into the atomic guiding region presents a hurdle to true CW operation of the device. To overcome this, in a first attempt we will pump atoms from the $|F=2,m_F=2>$ MOT ground state into the $|F=1,m_F=-1>$ dark state. This method has the unfortunate side effect of discarding $2/3$ of the atoms collected by the secondary MOT. As a better method, we envision MEMs-based light shutters integrated on-chip to block stray MOT light. In this mode, the guide would be loaded in a quasi-continuous fashion with cold-atom bunches from the secondary MOT at a rate of one to ten pulses per second. Once on the chip, the bunches would rapidly merge into a continuous flow. Since the light shutters would be synchronized with the MOT and atom transfer cycle in a way that no light reaches the chip, both levels $|F=2,m_F=2>$ and $|F=1,m_F=-1>$ could be used for magnetic guiding. Replacing the Zeeman slower with another chip containing an array of MOTs~\cite{Yun:motarray,trupke:motarray,jollenbeck:hexapolechip} would also provide benefit: a Zeeman slower for $^{87}\textrm{Rb}$ is a large and unwieldy device, not well suited to packaging for commercial applications.

\acknowledgements{This work was supported by NGIA grant \# HM1582-08-1-0028 and ASOFR grant \# FA9550-10-1-0371. The authors wish to thank Katsuo Kurabayahi for useful discussions.}

\bibliographystyle{h-physrev}
\bibliography{spiralguidersiv1a}

\end{document}